\documentclass[]{emulateapj}

\newcommand{\actaa}{Acta Astron.}

\let\gsim=\ga

\slugcomment{Accepted for publication in the Astrophysical Journal}
\shorttitle{CO Observations of the Host Galaxy of GRB~000418}
\shortauthors{Hatsukade et al.}

\begin{document}

\title{CO Observations of the Host Galaxy of GRB~000418 at $z = 1.1$}

\author{B. Hatsukade,\altaffilmark{1}
 		K. Kohno,\altaffilmark{2,3}
 		A. Endo,\altaffilmark{4}
 		K. Nakanishi,\altaffilmark{5}
 		and K. Ohta\altaffilmark{1}
 		}

\altaffiltext{1}{Department of Astronomy, Kyoto University, Kyoto 606-8502}
\altaffiltext{2}{Institute of Astronomy, the University of Tokyo, 2-21-1 Osawa, Mitaka, Tokyo 181-0015}
\altaffiltext{3}{Research Center for the Early Universe, University of Tokyo, 7-3-1 Hongo, Bunkyo, Tokyo 113-0033, Japan}
\altaffiltext{4}{Kavli Institute of NanoScience, Faculty of Applied Sciences, Delft University of Technology, Lorentzweg 1, 2628 CJ Delft, The Netherlands}
\altaffiltext{5}{National Astronomical Observatory of Japan, 2-21-1 Osawa, Mitaka, Tokyo 181-8588}

\begin{abstract}
We performed CO~($J = 2$--1) observations of the host galaxy of GRB~000418 at $z=1.1181$ with the Plateau de Bure Interferometer. 
Previous studies show that the host galaxy has properties similar to those of an ultraluminous infrared galaxy (ULIRG). 
The star-formation rate (SFR) of the host galaxy as derived from submillimeter and radio continuum emission is a few 100~$M_{\odot}$~yr$^{-1}$, which is an order of magnitude greater than the SFR derived from optical line emission. 
The large discrepancy between the SFRs derived from different observing wavelengths indicates the presence of a bulk of dust-obscured star formation and molecular gas that is enough to sustain the intense star formation. 
We failed to detect CO emission and derived 2$\sigma$ upper limits on 
the velocity integrated CO~(2--1) luminosity of $L'_{\rm CO} < 6.9 \times 10^9$~K~km~s$^{-1}$~pc$^2$ 
and the molecular gas mass of $M_{\rm H_2} < 5.5 \times 10^9$~$M_{\odot}$ 
by adopting a velocity width of 300~km~s$^{-1}$ 
and a CO-to-H$_2$ conversion factor of $\alpha_{\rm CO} = 0.8$~$M_{\odot}$~(K~km~s$^{-1}$~pc$^2$)$^{-1}$, which are standard values for ULIRGs. 
The lower limit on the ratio of far-infrared luminosity to CO luminosity, a measure of the star-formation efficiency, is higher compared to that of other gamma-ray burst hosts and other galaxy populations, which is consistent with active star formation taking place in this galaxy. 
\end{abstract}


\section{Introduction}

Long-duration gamma-ray bursts (GRBs)---the most energetic events in the universe---are considered to be effective indicators for star-formation activity in the distant universe \citep[e.g.,][]{tota97, wije98} because 
(1) GRBs are considered to occur because of the deaths of massive stars, and therefore, they are closely associated with star formation in host galaxies \citep[e.g.,][]{stan03, hjor03}, 
and (2) GRBs can be detected at cosmological distances \citep[the current record is $z = 8.2$ for GRB~090423;][]{tanv09, salv09}. 
The majority of GRB hosts are blue, subluminous, low-metal, dwarf star-forming galaxies \citep[e.g.,][]{fynb03, lefl03, fruc06} and their star-formation rates (SFRs) as derived from UV/optical/near-infrared (NIR) observations are typically $\sim$0.1--10~$M_{\odot}$~yr$^{-1}$ \citep[e.g.,][]{sava09, leve10, sven10}. 
However, if the production rate of GRBs traces the star formation in their hosts, a large fraction of GRBs should occur in star-forming regions obscured by dust. 
There are reports that some GRB hosts have a large amount of dust and obscured star formation: 
(1) There is a discrepancy between the SFRs derived from different observing wavelengths. 
SFRs derived from mid-infrared, submillimeter (submm), and radio wavelengths are much larger (by an order of magnitude in some cases) than SFRs derived from UV, optical, NIR wavelengths \citep[e.g.,][]{berg03, lefl06}. 
(2) Large hydrogen column densities ($N_{\rm H} \gsim 10^{22}$~cm$^{-2}$) are observed along the line of sight to GRBs \citep[e.g.,][]{jako06, scha07, zhen09}. 
(3) About 25--40\% of GRBs are ``dark GRBs'' \citep[e.g.,][]{fynb01, djor01, fynb09, grei11}.
The nature of dark GRBs, which are characterized by the faintness of their optical afterglow compared to their X-ray afterglow \citep{jako04, vand09}, is not yet well understood and one possible explanation is due to the large dust extinction along the line of sight to GRBs \citep[e.g.,][]{perl09}. 
So far, only a small fraction of GRB hosts have been studied for which high obscured star formation is indicated \citep[e.g.,][]{tanv04, prid06}. 
Whether GRB hosts have obscured star formation is still uncertain because it is difficult to identify them when their optical afterglows are extincted by dust.

An alternative approach for understanding star-formation activity in GRB hosts is to measure the amount of molecular gas, which is the ingredient for star formation. 
The CO emission line observations provide the information of molecular gas mass, dynamical mass, and star-formation efficiency in GRB hosts without being affected by dust extinction. 
Thus far, only a few efforts have been made to search for molecular gas in GRB hosts (Table~\ref{tab:summary}): 
CO~(1--0) observations of the GRB~030329 host \citep{kohn05, endo07}, 
CO~(3--2) observations of the GRB~980425 host \citep{hats07}, and 
CO~(3--2) observations of the GRB~090423 host \citep{stan11}. 
No CO emission has been detected from GRB hosts and whether GRB hosts have sufficient molecular gas to maintain their star formation remains unknown.

In this paper, we report a search for CO line emission toward the host galaxy of GRB~000418 using the Plateau de Bure Interferometer \citep[PdBI;][]{guil92}. 
\S~\ref{sec:host} describes the host galaxy of GRB~000418. 
\S~\ref{sec:observation} outlines the observations and data reduction, and the results are presented in \S~\ref{sec:result}. 
In \S~\ref{sec:discussion}, we derive constraints on physical quantities of the host galaxy and discuss the nature of the galaxy. 
A summary is presented in \S~\ref{sec:summary}.

Throughout the paper, we adopt a cosmology with $H_0=70$ km~s$^{-1}$ Mpc$^{-1}$, $\Omega_{\rm{M}}=0.3$, and $\Omega_{\Lambda}=0.7$.

\begin{deluxetable*}{ccccll}
\tabletypesize{\scriptsize}
\tablewidth{0pt}
\tablecaption{Summary of CO Observations in GRB Host Galaxies \label{tab:summary}}
\tablehead{
\colhead{GRB} & \colhead{$z$} & \colhead{Transition} & \colhead{$L'_{\rm CO}$} & \colhead{$M_{\rm H_2}$} & \colhead{Reference} \\
 & & & \colhead{(K~km~s$^{-1}$~pc$^2$)} & \colhead{($M_{\odot}$)} & 
} 
\startdata
GRB~980425 & 0.0085 & CO~(3--2) & $< 2.3 \times 10^7$ & $< 1.8 \times 10^8$ $^a$ & \cite{hats07} \\
GRB~000418 & 1.1181 & CO~(2--1) & $< 6.9 \times 10^9$ & $< 5.5 \times 10^9$ & {\it This work} \\
GRB~030329 & 0.1685 & CO~(1--0) & $< 4.6 \times 10^8$ & $< 1.9 \times 10^{10}$ $^b$ & \cite{kohn05, endo07} \\
GRB~090423 & 8.23   & CO~(3--2) & $< 3.6 \times 10^9$ & $< 2.9 \times 10^9$ $^c$ & \cite{stan11} 
\enddata
\tablecomments{
Upper limits are 2$\sigma$. \\
$^a$$\alpha_{\rm CO} = 8.0$~$M_{\odot}$~(K~km~s$^{-1}$~pc$^2$)$^{-1}$ is adopted. 
$^b$$\alpha_{\rm CO} = 40$~$M_{\odot}$~(K~km~s$^{-1}$~pc$^2$)$^{-1}$ is adopted. 
$^c$$\alpha_{\rm CO} = 0.8$~$M_{\odot}$~(K~km~s$^{-1}$~pc$^2$)$^{-1}$ is adopted. 
}
\end{deluxetable*}

\section{Host Galaxy of GRB~000418}\label{sec:host}

UV/optical/NIR observations show that the host galaxy is a blue, compact, subluminous galaxy \citep[$M_B = -20.6$;][]{goro03} at $z=1.1181 \pm 0.0001$ \citep{bloo03}. 
The extinction-corrected SFR derived from the [O\,{\sc ii}] line luminosity is 15.4~$M_{\odot}$~yr$^{-1}$ \citep{bloo03, goro03}. 
SED fits to the UV/optical/NIR data show that the host is a young star-forming galaxy with an SFR of $\sim$10--20~$M_{\odot}$~yr$^{-1}$ \citep{goro03, chri04, sava09, sven10}. 
Submm and radio observations with the Submillimetre Common-User Bolometer Array (SCUBA) and the Very Large Array (VLA) detected a source at the position of the host galaxy with fluxes of 
$S(\rm 850\mu m) = 3.2 \pm 0.9$~mJy, 
$S(\rm 1.43GHz) = 69\pm 15$~$\mu$Jy, 
$S(\rm 4.86GHz) = 46 \pm 13$~$\mu$Jy, 
and $S(\rm 8.46GHz) = 51 \pm 12$~$\mu$Jy \citep{berg03}. 
The SFRs derived from the submm and radio emissions are ${\rm SFR(submm)} = 690 \pm 195$~$M_{\odot}$~yr$^{-1}$ and ${\rm SFR(radio)} = 330 \pm 75$~$M_{\odot}$~yr$^{-1}$ \citep{berg03}. 
The SED fit of \cite{mich08} ranging from UV to radio wavelengths shows that the host galaxy is a young star-forming galaxy with $L_{\rm IR} = 4.6 \times 10^{12}$~$L_{\odot}$ and ${\rm SFR} = 288$~$M_{\odot}$~yr$^{-1}$. 
The IR luminosity classifies the host galaxy as an ultraluminous infrared galaxy (ULIRG). 
The large discrepancy between the SFR based on UV/optical/NIR observations and the SFR based on submm/radio observations indicates that the bulk of the star formation is obscured by dust.

\section{Observations and Data Reduction}\label{sec:observation}

The PdBI observations were conducted on August 6 and 7, 2006 using the D configuration with five antennas and on March 13, 2007 using the B configuration with six antennas. 
The phase center was positioned at $\alpha$(J2000)~=~$12^h$~$25^m$~$19.3^s$ and $\delta$(J2000)~=~$+20^{\circ}$~$06'$~$11''.0$. 
The redshifted CO~(2--1) and CO~(4--3) lines were simultaneously observed at 3-mm and 1.3-mm bands, respectively. 
Receiver~1 was tuned to 108.842~GHz (3-mm band) for the upper sideband and the receiver~2 was tuned to 217.667~GHz (1.3-mm band) for the lower sideband. 
The correlator was equipped with 580-MHz bandwidth in each sideband in the 2006 observations and 1-GHz bandwidth in each sideband in the 2007 observations. 
The system temperature of receiver~1 was $T_{\rm sys} \sim 200$--300~K in 2006 and $T_{\rm sys} \sim 100$--200~K in 2007 (SSB). 
Because the atmospheric conditions in the 1.3-mm band for the CO~(4--3) line were unfavorable ($T_{\rm sys} \gsim 1000$~K), we use only 3-mm data for the CO~(2--1) line in what follows.

Data reduction and imaging were carried out using the CLIC program in the GILDAS package \citep{guil00}. 
Passband calibrations were performed using bright QSOs observed during the track. 
Flux calibrations were performed using standard calibrators. 
Maximum sensitivity was achieved by adopting natural weighting, which gave a final synthesized beam size of 
$2\farcs54 \times 1\farcs56$ (position angle = $13^{\circ}$).

\section{Results}\label{sec:result}

\begin{figure}[t]
\begin{center}
\includegraphics[width=\linewidth]{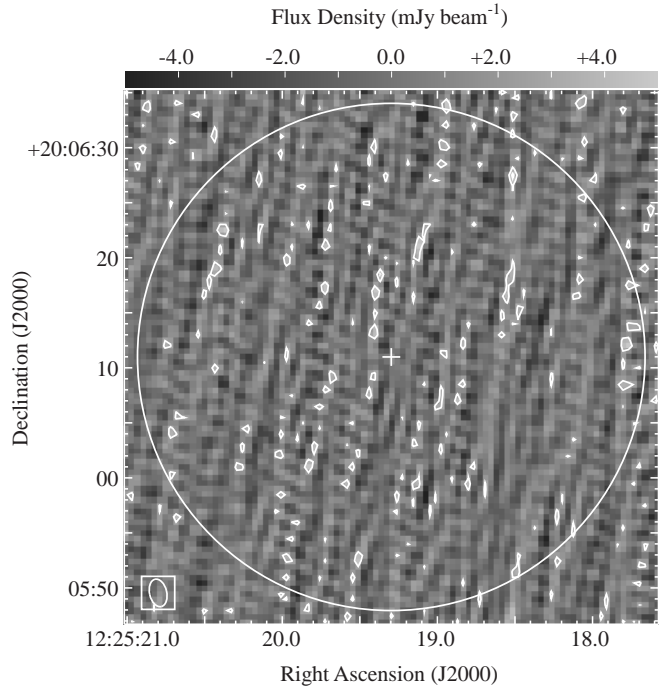}
\end{center}
\caption{
Map around the position of GRB~000418 (cross) at 108.83 GHz with a velocity resolution of 100~km~s$^{-1}$ (36.3~MHz). 
Contours are $+2$$\sigma$ (1$\sigma$ = 1.3~mJy~beam$^{-1}$). 
The large circle represents the field of view (46$''$ diameter). 
The synthesized beam is shown on the bottom left. 
}
\label{fig:map}
\end{figure}

Neither CO line emission nor continuum emission is detected (Fig.~\ref{fig:map}). 
The rms noise level is 1.3~mJy~beam$^{-1}$ with 100~km~s$^{-1}$ resolution. 
Summing the signals within the bandwidth, we obtained an rms noise level of 0.15~mJy~beam$^{-1}$ for the 3-mm continuum (rest frame 1.3~mm). 
The 2$\sigma$ upper limits on the CO flux and continuum flux at the position of the host are 2.4~mJy~beam$^{-1}$ (100~km~s$^{-1}$ resolution) and 0.62~mJy~beam$^{-1}$, respectively. 
This upper limit on continuum flux is consistent with the SED model of \cite{mich08}. 

The CO line luminosity ($L'_{\rm CO}$) is given as follows \citep{solo92}: 
\begin{eqnarray}
L'_{\rm CO}= 3.25 \times 10^7 S_{\rm CO} \Delta v \nu_{\rm obs}^{-2} D_L^2 (1+z)^{-3},
\end{eqnarray}
where $L'_{\rm CO}$ is measured in K~km~s$^{-1}$~pc$^2$, $S_{\rm CO}$ is the observed CO flux in Jy, $\Delta v$ is the velocity width in km~s$^{-1}$, and $D_L$ is the luminosity distance in Mpc. 
Assuming a velocity width of $300\ \rm{km\ s^{-1}}$, which is the typical value for local ULIRGs \citep{solo97}, the 2$\sigma$ upper limit of CO~(2--1) line luminosity is $L'_{\rm CO}$(2--1) $< 6.9 \times 10^9$~K~km~s$^{-1}$~pc$^2$.

\section{Discussion}\label{sec:discussion}

\subsection{Constraints on Physical Quantities}

\subsubsection{Molecular Gas Mass}\label{sec:gas-mass}
Molecular gas mass is given by
\begin{eqnarray}
M_{\rm H_2} = \alpha_{\rm CO} L'_{\rm CO}, 
\end{eqnarray}
where $\alpha_{\rm CO}$ is the CO-to-H$_2$ conversion factor in $M_{\odot}$~(K~km~s$^{-1}$~pc$^2$)$^{-1}$. 
We assume that the gas is optically thick and thermalized and has a CO~(2--1)/CO~(1--0) luminosity ratio of unity. 
The 2$\sigma$ upper limit on molecular gas mass is $M_{\rm H_2} < 5.5 \times 10^9$~$M_{\odot}$, and it is obtained by adopting a conversion factor of $\alpha_{\rm CO} = 0.8$~$M_{\odot}$~(K~km~s$^{-1}$~pc$^2$)$^{-1}$, which is the standard value for ULIRGs \citep{down98}. 
This is lower than the median value of $(3.0 \pm 1.6) \times 10^{10}$~$M_{\odot}$ obtained for a sample of submillimeter galaxies \citep[SMGs;][]{grev05}. 
Note that if we adopt a Galactic conversion factor of $\alpha_{\rm CO} = 4.6$~$M_{\odot}$~(K~km~s$^{-1}$~pc$^2$)$^{-1}$ \citep{solo91}, the 2$\sigma$ upper limit would increase by about a factor of 6.

\subsubsection{Dust Mass}\label{sec:dust-mass}
The dust mass can be derived from submm flux, 
\begin{eqnarray}
M_d = \frac{S_{\rm obs}D_L^2}{(1+z)\kappa_d(\nu_{\rm rest})B(\nu_{\rm rest}, T_d)},
\end{eqnarray}
where $S_{\rm obs}$ is the observed flux density, $\nu_{\rm rest}$ is the rest frequency, $\kappa_d(\nu_{\rm rest})$ is the rest-frequency dust mass absorption coefficient, $T_d$ is the dust temperature, and $B(\nu_{\rm rest}, T_d)$ is the Planck blackbody function \citep{hugh97}. 
It is believed that the absorption coefficient varies with frequency as $\kappa_d \propto \nu^{\beta}$, where $\beta$ lies between 1 and 2 \citep[e.g,][]{hild83}. 
We assume $\kappa_d(125{\rm \mu m}) = 1.875$~m$^2$~kg$^{-1}$ \citep{hild83}.
$T_d$ and $\beta$ depend on the properties of the dust, and hence, on the type of galaxy being considered. 
Because the submm flux is detected only at 850~$\mu$m \citep{berg03}, $T_d$ and $\beta$ cannot be uniquely determined. 
We adopt $T_d = 30$--50~K and $\beta = 1.5$, the typical values for local ULIRGs and SMGs \citep[e.g.,][]{yang07, kova06, copp08, mich10}. 
The dust mass is calculated to be $M_d = (4$--$10) \times 10^8$~$M_{\odot}$. 
This is consistent with $M_d = 8.2 \times 10^8$~$M_{\odot}$ derived from SED model fit of \cite{mich08}, where they assume $T_d = 50$~K.

From the 2$\sigma$ upper limit of molecular gas mass derived in \S~\ref{sec:gas-mass}, the 2$\sigma$ upper limit of molecular gas-to-dust mass ratio is estimated to be $\sim$10--20. 
This is lower than that of other galaxy samples, such as 
100--200 of our Galaxy \citep[e.g.,][]{hild83} and $\sim$50--a few 100 for spiral galaxies and star-forming galaxies at local to high-redshift \citep[e.g.,][]{dunn00, stev05, kova06, mich10}. 
We note that the derived molecular gas-to-dust ratio relies on the estimates of both gas mass and dust mass that are themselves quite uncertain. 
If we adopt a Galactic CO-to-H$_2$ conversion factor for deriving molecular gas mass, the upper limit on molecular gas-to-dust ratio would increase by about a factor of 6.

\subsubsection{Star-formation Efficiency}\label{sec:sfe}

The CO luminosity and the far-infrared (FIR) luminosity are measures of the molecular gas mass and SFR, respectively.
Therefore, the ratio of $L_{\rm{FIR}}/L'_{\rm{CO}}$ indicates how efficiently stars are formed from molecular gas and is used as an indicator of star-formation efficiency \citep[SFE;][]{youn86}. 
The 2$\sigma$ lower limit is $L_{\rm{FIR}}/L'_{\rm{CO}} > 6.7 \times 10^2$~$L_{\odot}$~(K~km~s$^{-1}$~pc$^2$)$^{-1}$ and is obtained using the FIR luminosity derived by \cite{mich08}. 
This is higher than that of local spiral galaxies \citep[$\sim$10--100;][]{youn96}, LIRGs, and ULIRGs \citep[$\sim$a few hundred;][]{sand91, solo97} and is located at the higher end of SMGs and QSOs \citep[see][and references therein]{solo05}, indicating that active star formation occurs in the host galaxy. 
This is shown in terms of a specific star formation rate (SSFR; SFR per unit stellar mass). 
SSFR is considered to be an indicator of current star-forming activity, and its inverse is related to the mass doubling time.
The SSFR of 12--15~Gyr$^{-1}$ derived in previous studies \citep{chri04, mich08, sven10} is higher than that of other galaxy populations in the local to high-redshift universe \citep[e.g.,][]{cast06}. 
It is known that GRB hosts have higher SSFRs compared to field galaxies \citep[e.g.,][]{char02, chri04, cast06, sava09}. 
The high SFE of this host galaxy is consistent with the typical properties of GRB hosts. 

The high SFE could be due to the uncertainty of the CO luminosity derived in this work and/or the FIR luminosity derived from submm emission, and we discuss this issue in the next section.

\subsection{Nondetection of CO}

There are some possible reasons for the nondetection of CO emission. 
One is that the amount of molecular gas in this galaxy is actually small. 
Optical observations show that the galaxy is compact and subluminous \citep{bloo03}. 
However, in order to sustain the large SFR inferred from submm and radio observations, a large amount of molecular gas is essential. 
It is possible that the submm flux is overestimated. 
When dealing with a low S/N map, we must account for the boosting of flux densities of low S/N sources to above detection thresholds \citep{murd73, hogg98}. 
In surveys at submm wavelengths, flux densities of low S/N sources ($\sim$3.5$\sigma$) are corrected downward by $\sim$10--30\% \citep[e.g.,][]{copp06, hats11}. 
In addition, some of the detections in flux-limited submm surveys are spurious, having been caused by positive noise fluctuations. 
Because the submm observations of \cite{berg03}, where S/N $\sim$ 3.5 for the submm source, do not take into account these effects, the submm derived SFR is possibly overestimated. 
It is also possible that the radio flux is overestimated. 
\cite{berg03} pointed out that the contribution from the afterglow to the radio flux is expected to be 10, 5, and 10~$\mu$Jy at 1.43, 4.86, and 8.46~GHz, respectively, suggesting that the intrinsic radio flux of the host galaxy could decrease by about 10--20\%.

Another possibility is that the CO emission per unit molecular gas mass is low. 
Some authors suggest that the CO-to-H$_2$ conversion factor depends on the oxygen abundance; 
the CO-to-H$_2$ conversion factor increases as the metallicity of the host galaxy decreases \citep[e.g.,][]{wils95, arim96}. 
CO observations of local low-metal dwarf galaxies suggest a higher CO-to-H$_2$ conversion factor \citep[e.g.,][]{tayl98, lero05, komu11}. 
GRB hosts typically have subsolar metallicity \citep[e.g.,][]{fynb03, stan06, leve10}, and theoretical models support the low-metal environment of GRB progenitors \citep[e.g.,][]{woos06}. 
The metallicity of the host galaxy of GRB~000418, $[12+\log(\rm O/H)] = 8.43$, was derived by \cite{sven10} based on the stellar mass-metallicity relation of \cite{sava05}. 
The correlation between $\alpha_{\rm CO}$ and metallicity obtained by \cite{arim96} yields $\alpha_{\rm CO} = 12$~$M_{\odot}$~(K~km~s$^{-1}$~pc$^2$)$^{-1}$. 
This is 15 times higher than that of the ULIRGs we adopt in this work. 
Therefore, it is possible that the nondetection of CO is due to the low metallicity of the host galaxy.

\section{Summary}\label{sec:summary}

We carried out observations of redshifted CO~($J = 2$--1) line toward the host galaxy of GRB~000418 at $z=1.1181$ using the PdBI. 
Neither CO line emission nor continuum emission were detected. 
We derived 2$\sigma$ upper limits on 
the velocity integrated CO luminosity $L'_{\rm CO} < 6.9 \times 10^9$~K~km~s$^{-1}$~pc$^2$ 
and the molecular gas mass $M_{\rm H_2} < 5.5 \times 10^9$~$M_{\odot}$, 
by assuming a velocity width of $300\ \rm{km\ s^{-1}}$ 
and a CO-to-H$_2$ conversion factor of 0.8~$M_{\odot}$~(K~km~s$^{-1}$~pc$^2$)$^{-1}$, which are standard values for local ULIRGs. 
The upper limit of molecular gas-to-dust mass ratio was lower than that of other galaxy populations, although it must be noted that the derived molecular gas mass and dust mass were uncertain. 
The lower limit on the ratio of FIR luminosity to CO luminosity, a measure of the star-formation efficiency, was higher compared to that of other GRB hosts and other galaxy populations, indicating that active star formation is taking place in this galaxy. \\

No CO line emission has been detected in GRB hosts so far. 
Previous studies show that a CO detection itself is difficult in low-metal systems. 
We expect that the Atacama Large Millimeter/submillimeter Array (ALMA) will enable us to detect CO lines in such challenging conditions and to examine gas properties of GRB hosts.

\acknowledgments

We would like to acknowledge P.~Salome, R.~Neri, and the IRAM staff for the help they provided during the observation and data reduction. 
We are grateful to T.~Okuda, F.~Egusa, and Y.~Tamura for their help in data reduction and useful suggestions. 
We thank the referee for helpful comments and suggestions. 
This work was supported by the Grant-in-Aid for the Global COE Program ``The Next Generation of Physics, Spun from Universality and Emergence'' from the Ministry of Education, Culture, Sports, Science and Technology (MEXT) of Japan. 
This work was partly supported by the Foundation for Promotion of Astronomy. 
B.H. is supported by Research Fellowship for Young Scientists from the Japan Society of the Promotion of Science (JSPS). 
A.E. is financially supported by NWO (Veni grant 639.041.023) and JSPS Fellowship for Research Abroad.


\end{document}